\documentclass[12pt]{article}

\usepackage{amsmath}
\usepackage{amssymb}
\usepackage{amsthm}
\usepackage[pdfpagemode=None]{hyperref}
\usepackage{graphicx}
\usepackage{pstricks}

\newtheorem{thm}{Theorem}[section]

\newtheorem{lem}[thm]{Lemma}

\newtheorem{assumption}[thm]{Assumption}

\newtheorem{definition}[thm]{Definition}

\newtheorem{example}[thm]{Example}

\newtheorem{remark}[thm]{Remark}
\newenvironment{rem}{\begin{remark}\rm}{\end{remark}}
\newtheorem{tab}{Table}

\newcommand{\diag}{{\rm diag}\,}

\newcommand{\High}{{\rm High}\,}
\newcommand{\Low}{{\rm  Low}\,}

\def\eps{\varepsilon}

\title{Derivatives of Entropy Rate in Special Families of Hidden Markov
Chains}

\author{Guangyue Han, Brian Marcus\\
  {\normalsize Department of Mathematics}\vspace{-1mm} \\
  {\normalsize University of British Columbia}\vspace{-1mm} \\
  {\normalsize Vancouver, B.C.,  V6T 1Z2}\\
  {\normalsize {\em e-mail:\/} {ghan, marcus}@math.ubc.ca}}

\date{{\normalsize \today}}

\begin{document}\maketitle\thispagestyle{empty}


\begin{abstract}
Consider a hidden Markov chain obtained as the observation process
of an ordinary Markov chain corrupted by noise. Zuk, et.
al.~\cite{zu04, zu05} showed how, in principle, one can explicitly
compute the derivatives of the entropy rate of at extreme values of
the noise. Namely, they showed that the derivatives of standard
upper approximations to the entropy rate actually stabilize at an
explicit finite time.  We generalize this result to a natural class
of hidden Markov chains called ``Black Holes.''  We also discuss in
depth special cases of binary Markov chains observed in binary
symmetric noise, and give an abstract formula for the first
derivative in terms of a measure on the simplex due to Blackwell.
\end{abstract}

\section{Introduction}

As in~\cite{gm05}, let $Y=\{Y_{-\infty}^{\infty}\}$ be a stationary
Markov chain with a finite state alphabet $\{1, 2, \cdots, B\}$. A
function $Z=\{Z_{-\infty}^{\infty}\}$ of the Markov chain $Y$ with
the form $Z=\Phi(Y)$ is called a hidden Markov chain; here $\Phi$ is
a finite valued function defined on $\{1, 2, \cdots, B\}$, taking
values in $\{1, 2, \cdots, A\}$. Let $\Delta$ denote the probability
transition matrix for $Y$; it is well known that the entropy rate
$H(Y)$ of $Y$ can be analytically expressed using the stationary
vector of $Y$ and $\Delta$. Let $W$ be the simplex, comprising the
vectors
$$
\{w=(w_1, w_2, \cdots, w_B) \in \mathbb{R}^B:w_i \geq 0, \sum_i
w_i=1\},
$$
and let $W_a$ be all $w \in W$ with $w_i=0$ for $\Phi(i) \neq a$.
For $a \in A$, let $\Delta_a$ denote the $B \times B$ matrix such
that $\Delta_a(i, j)=\Delta(i, j)$ for $j$ with $\Phi(j)=a$, and
$\Delta_a(i, j)=0$ otherwise. For $a \in A$, define the
scalar-valued and vector-valued functions $r_a$ and $f_a$ on $W$ by
$$
r_a(w)= w \Delta_a \mathbf{1},
$$
and
$$
f_a(w)=w \Delta_a/ r_a(w).
$$
Note that $f_a$ defines the action of the matrix $\Delta_a$ on the
simplex $W$.

If $Y$ is irreducible, it turns out that
\begin{equation}
\label{blackwell_form} H(Z)=-\int \sum_a r_a(w) \log r_a(w) dQ(w),
\end{equation}
where $Q$ is {\em Blackwell's measure}~\cite{bl57} on $W$.  This measure is
defined as the limiting distribution $p(y_0 = \cdot
|z_{-\infty}^0)$.


Recently there has been a great deal of work on the entropy rate of
a hidden Markov chain~\cite{or03, ho03, or04, zu04, ja04,zu05}. See
also closely related work~\cite{ma84a1, pe90, pe92}.


In Section~\ref{stabilizing}, we establish a ``stabilizing''
property for the derivatives of the entropy rate in a family we call
``Black Holes''. Using this property, one can, in principle,
explicitly calculate the derivatives of the entropy rate for this
case.

In Section~\ref{particular} we consider binary Markov chains
corrupted by binary symmetric noise. For this class, we obtain
results on the support of Blackwell's measure, and for a special
case, that we call the ``non-overlapping'' case, we express the
first derivative of the entropy rate as the sum of terms, involving
Blackwell's measure, which have meaningful interpretations. We also
show how this expression relates to earlier examples, given
in~\cite{or04}, of non-smoothness on the boundary for this class of
hidden Markov chains, and we compute the second derivative in an
important special case.

\section{Stabilizing Property of Derivatives in Black Hole Case}
\label{stabilizing}

Suppose that for every $a \in A$, $\Delta_a$ is a rank one matrix,
and every column of $\Delta_a$ is either strictly positive or all
zeros. In this case, the image of $f_a$ is a single point and each
$f_a$ is defined on the whole simplex $W$. Thus we call this  the
{\em Black Hole} case.  Analyticity of the entropy rate at a Black
Hole follows from Theorem 1.1 of~\cite{gm05}.

In this section we show that, in principle, the coefficients of a
Taylor series expansion, centered at a Black Hole, can be explicitly
computed. This result was motivated by and generalizes earlier work
by Zuk, et. al.~\cite{zu04, zu05} and
Ordentlich-Weissman~\cite{or05} on cases of hidden Markov chains
obtained by passing a Markov chain through special kinds of
channels. All of the hidden Markov chains considered in~\cite{zu04,
zu05} are Black Holes.

As an example, consider a hidden Markov chain obtained from a binary
Markov chain corrupted by binary symmetric noise with crossover
probability $\eps$ (described in Example 4.1 of~\cite{gm05}).  When
$\eps =0$,
$$
\Delta = \left [ \begin{array}{cccc}
\pi_{00}  & 0 & \pi_{01}  & 0\\
\pi_{00} & 0 & \pi_{01}  & 0\\
\pi_{10}  & 0 & \pi_{11}  & 0\\
\pi_{10}  & 0 & \pi_{11}  & 0\\
\end{array} \right ];
$$
here, $\pi_{ij}$'s are the Markov transition probabilities, and
$\Phi$ maps states 1 and 4 to 0 and maps states 2 and 3 to 1. In
this case, the nonzero entries of $\Delta_0$ and $\Delta_1$ are
restricted to a single column and so both $\Delta_0$ and $\Delta_1$
have rank one.  If $\pi_{ij}$'s are all positive, then this is a
Black Hole case.
\bigskip

Suppose that $\Delta$ is analytically parameterized by a vector
variable $\varepsilon=(\varepsilon_1, \varepsilon_2, \cdots,
\varepsilon_m)$. Recall that $H_n(Z)$ is defined as
$$
H_n(Z)=H(Z_0|Z_{-n}^{-1}).
$$
The following theorem says that at a Black Hole, one can calculate
the derivatives of $H(Z)$ by taking the derivatives of $H_n(Z)$ for
large enough $n$.

\begin{thm}  \label{stabilizingtheorem1}
If at $\varepsilon=\hat{\varepsilon}$, for every $a \in A$,
$\Delta_a$ is a rank one matrix, and every column of $\Delta_a$ is
either a positive or a zero column, then
$$
\hspace{-5mm} \left.
\frac{\partial^{\alpha_1+\alpha_2+\cdots+\alpha_m}
H(Z)}{\partial_{\varepsilon_1}^{\alpha_1}
\partial_{\varepsilon_2}^{\alpha_2} \cdots
\partial_{\varepsilon_m}^{\alpha_m}} \right |_{\varepsilon=\hat{\varepsilon}}=\left. \frac{\partial^{\alpha_1+\alpha_2+\cdots+\alpha_m}
H_{\alpha_1+\alpha_2+\cdots+\alpha_m}(Z)}{\partial_{\varepsilon_1}^{\alpha_1}
\partial_{\varepsilon_2}^{\alpha_2} \cdots
\partial_{\varepsilon_m}^{\alpha_m}} \right |_{\varepsilon=\hat{\varepsilon}}
$$
\end{thm}
In fact, we give a stronger result,
Theorem~\ref{stabilizingtheorem2}, later in this section.
\begin{proof}
For simplicity we assume that $\Delta$ is only parameterized by one
real variable $\varepsilon$, and we drop $\varepsilon$ when the
implication is clear from the context.

We shall first prove that for all sequences $z_{-\infty}^{0}$ the
$n$-th derivative of $p(z_0|z_{-\infty}^{-1})$ stabilizes:
\begin{equation}   \label{conditional}
p^{(n)}(z_0|z_{-\infty}^{-1})=p^{(n)}(z_0|z_{-n-1}^{-1})~~\mbox{at }
\varepsilon=\hat{\varepsilon}.
\end{equation}

Since $p(z_0|z_{-\infty}^{-1})=p(y_{-1}=\cdot \;|z^{-1}_{-\infty})
\Delta_{z_0} \mathbf{1}$ (here $\cdot$ represent the states of the
Markov chain $Y$), it suffices to prove that for the $n$-th
derivative of $x_i=p(y_i=\cdot \;|z^i_{-\infty})$, we have
  \begin{equation}   \label{conditional2}
x_i^{(n)}=p^{(n)}(y_i=\cdot \;|z_{-\infty}^i)=p^{(n)}(y_i=\cdot
\;|z_{i-n}^i) \mbox{ at } \varepsilon=\hat{\varepsilon}.
\end{equation}

Consider the iteration:
$$
x_i=\frac{x_{i-1} \Delta_{z_i}}{x_{i-1} \Delta_{z_i} \mathbf{1}}.
$$
In other words, $x_i$ can be viewed as a function of $x_{i-1}$ and
$\Delta_a$. Let $g$ denote this function. Since at
$\varepsilon=\hat{\varepsilon}$, $\Delta_{z_i}$ is a rank one
matrix, we conclude that $g$ is constant as a function of $x_{i-1}$.
Thus at $\varepsilon=\hat{\varepsilon}$
\begin{equation}  \label{zeroorder}
\hspace{-0.5cm} x_i=p(y_i=\cdot \;|z^i_{-\infty})=\frac{x_{i-1}
\Delta_{z_i}}{x_{i-1} \Delta_{z_i} \mathbf{1}}=\frac{p(y_{i-1}=\cdot
\;) \Delta_{z_i}}{p(y_{i-1}=\cdot \;) \Delta_{z_i}
\mathbf{1}}=p(y_i=\cdot \;|z_i).
\end{equation}

Taking the derivative of $g$ with respect to $\varepsilon$, we have
at $\varepsilon=\hat{\varepsilon}$,
$$
x_i'=\left. \frac{\partial g}{\partial \Delta_{z_i}}
\right|_{\varepsilon=\hat{\varepsilon}} (x_{i-1}, \Delta_{z_i}) \;
\Delta_{z_i}' +\left. \frac{\partial g}{\partial x_{i-1}}
\right|_{\varepsilon=\hat{\varepsilon}} (x_{i-1}, \Delta_{z_i}) \;
x_{i-1}'.
$$

Since at $\varepsilon=\hat{\varepsilon}$, $g$ is a constant as a
function of $x_{i-1}$, we have
$$
\left. \frac{\partial g}{\partial x_{i-1}}
\right|_{\varepsilon=\hat{\varepsilon}} (x_{i-1},\Delta_{z_i})
=\frac{\partial (\mbox{a constant vector})}{\partial x_{i-1}}=0.
$$
It then follows from (\ref{zeroorder}) that at
$\varepsilon=\hat{\varepsilon}$
$$
x'_i=p'(y_i=\cdot \;|z_{-\infty}^i)=p'(y_i=\cdot \;|z_{i-1}^i).
$$
Taking higher order derivatives, we have
$$
x_i^{(n)}=\left. \frac{\partial g}{\partial x_{i-1}}
\right|_{\varepsilon=\hat{\varepsilon}} (x_{i-1}, \Delta_{z_i}) \;
x_{i-1}^{(n)}+ \mbox{other terms},
$$
where ``other terms'' involve only lower order (than $n$)
derivatives of $x_{i-1}$. By induction, we conclude that
$$
x_i^{(n)}=p^{(n)}(y_i=\cdot \;|z_{-\infty}^i)=p^{(n)}(y_i=\cdot
\;|z_{i-n}^i).
$$
at $\varepsilon=\hat{\varepsilon}$. We then have
(\ref{conditional2}) and therefore (\ref{conditional}) as desired.

By the proof of Theorem 1.1 of~\cite{gm05}, the complexified
$H_n(Z)$ uniformly converges to the complexified $H(Z)$, and so  we
can switch the limit operation and the derivative operation. Thus,
at all $\varepsilon$,
$$
H'(Z)=(\lim_{k \to \infty} \sum_{z_{-k}^0} (p(z_{-k}^0) \log p(z_0|
z_{-k}^{-1}))'
$$
$$
=\lim_{k \to \infty} \sum_{z_{-k}^0} (p'(z_{-k}^0) \log p(z_0|
z_{-k}^{-1})+ p(z_{-k}^0) \frac{p'(z_0| z_{-k}^{-1})}{p(z_0|
z_{-k}^{-1})})
$$
Since
$$
\sum_{z_0} p(z_{-k}^0) \frac{p'(z_0| z_{-k}^{-1})}{p(z_0|
z_{-k}^{-1})}=\sum_{z_0} p(z_{-k}^{-1}) p'(z_0| z_{-k}^{-1})=0,
$$
we have for all $\varepsilon$
\begin{equation}  \label{firstorder}
H'(Z)=\lim_{k \to \infty} \sum_{z_{-k}^0} (p'(z_{-k}^0) \log p(z_0|
z_{-k}^{-1})).
\end{equation}

At $\varepsilon=\hat{\varepsilon}$, we obtain:
$$
H'(Z)=\lim_{k \to \infty} \sum_{z_{-k}^0} (p'(z_{-k}^0) \log
p(z_0|z_{-1}))
$$
$$
=\sum_{z_{-1}^0} (p'(z_{-1}^0) \log p(z_0|z_{-1}))=H'_1(Z).
$$
For higher order derivatives, again using the fact that we can
interchange the order of limit and derivative operations and using
(\ref{firstorder}) and Leibnitz formula, we have for all
$\varepsilon$
$$
H^{(n)}(Z)=\lim_{k \to \infty} \sum_{z_{-k}^0} \sum_{l=1}^n
C_{n-1}^{l-1} p^{(l)}(z_{-k}^0) (\log p(z_0|z^{-1}_{-k}))^{(n-l)}
$$
(the use of (\ref{firstorder}) accounts for the fact that there is
no $l =0$ term in this expression).  Note that the term $(\log
p(z_0|z^{-1}_{-k}))^{(n-l)}$ involves only the lower order (less
than or equal to $n-1$) derivatives of $p(z_0|z^{-1}_{-k})$, which
are already ``stabilizing'' in the sense of (\ref{conditional}); so,
we have
$$
\hspace{-0.2cm} H^{(n)}(Z)=\lim_{k \to \infty} \sum_{z_{-k}^0}
\sum_{l=1}^n C_{n-1}^{l-1} p^{(l)}(z_{-k}^0) (\log
p(z_0|z^{-1}_{-n}))^{(n-l)}
$$
$$
=\sum_{z_{-n}^0} \sum_{l=1}^n C_{n-1}^{l-1} p^{(l)}(z_{-n}^0) (\log
p(z_0|z^{-1}_{-n}))^{(n-l)}=H^{(n)}_n(Z).
$$
We thus prove the theorem.
\end{proof}

\begin{rem}
It follows from (\ref{zeroorder}) that a hidden Markov chain at a
Black Hole is, in fact, a Markov chain. Note that in the argument
above the proof of the stabilizing property of the first derivative
(as opposed to higher derivatives) requires only that the hidden
Markov chain is Markov and that we can interchange the order of
limit and derivative operations (instead of the stronger Black Hole
property). Therefore if a hidden Markov chain $Z$ defined by $\hat
\Delta$ and $\Phi$ is in fact a Markov chain, and the complexified
$H_n(Z)$ uniformly converges to $H(Z)$ on some neighborhood of $\hat
\Delta$ (e.g., if the conditions of Theorem 1.1, 6.1 or 7.5
of~\cite{gm05} hold), then  at $\hat \Delta$, we have
\begin{equation} \label{infactmarkov}
H'(Z)=H'_1(Z).
\end{equation}
For instance, consider the following hidden Markov chain $Z$ defined
by
$$
\hat{\Delta}=\left[ \begin{array}{ccc}
        1/4&1/4/&1/2\\
        0&1/6&5/6\\
        7/8&1/8&0\\
       \end{array} \right],
$$
with $\Phi(1)=0$ and $\Phi(1)=\Phi(2)=1$. $Z$ is in fact a Markov
chain (see page $134$ in \cite{ke60}), and one checks that
$\hat{\Delta}$ satisfies the conditions in Theorem 7.5
in~\cite{gm05}. We conclude that for this example,
(\ref{infactmarkov}) holds.
\end{rem}
\vspace{.3 in}

In the cases studied in ~\cite{zu04, zu05, or05}, the authors
obtained, using a finer analysis, a shorter ``stabilizing length.''
This shorter length can be derived for the Black Hole case as well,
as shown in Theorem~\ref{stabilizingtheorem2} below, even though the
proof in~\cite{zu05} doesn't seem to work.

We need some preliminary lemmas for the proof of
Theorem~\ref{stabilizingtheorem2}.

By induction, one can prove that the formal derivative of $y \log y$
takes the following form:
$$
(y \log y)^{(N)}=\sum_{a_1 \geq a_2 \geq \cdots \geq a_{m+1}:
a_1+a_2+\cdots+a_{m+1}=N} E_{[a_1, a_2, \cdots, a_{m+1}]}
\frac{y^{(a_1)}y^{(a_2)} \cdots y^{(a_{m+1})}}{y^m}+y^{(N)} (\log
y+1)
$$
$$
\hspace{-1cm}=\sum_{i=1}^{N-1} y^{(a_1=i)} \sum_{a_2 \geq a_3 \geq
\cdots \geq a_{m+1}} E_{[a_1, a_2, \cdots, a_{m+1}]}
\frac{y^{(a_2)}y^{(a_3)} \cdots y^{(a_{m+1})}}{y^m}+y^{(N)} (\log
y+1).
$$
Let $q_i[y]$ denote the ``coefficient'' of $y^{(i)}$, which is a
function of $y$ and its formal derivatives (up to the $i$-th order
derivative). Thus we have
$$
(y \log y)^{(N)}=\sum_{i=1}^N q_i[y] y^{(i)}=\High_N[y]+\Low_N[y],
$$
where $\High_N[y]=\sum_{\lceil (N+1)/2 \rceil}^N q_i[y] y^{(i)}$ and
$\Low_N[y]=\sum_{i=1}^{\lceil (N-1)/2 \rceil} q_i[y] y^{(i)}$.

In the following, let $P(a_1, a_2, \cdots, a_m)$ denote the number
of distinct sequences obtained by permuting the coordinates of the
sequence $(a_1, a_2, \cdots, a_m)$. Namely if
\begin{equation}   \label{aaa}
\hspace{-0.8cm} a_1=a_2=\cdots=a_{m_1}
>a_{m_1+1}=\cdots=a_{m_1+m_2}> \cdots >
a_{m_1+m_2+\cdots+m_{j-1}+1}=\cdots=a_{m_1+m_2+\cdots+m_j}=a_m,
\end{equation}
then
$$
P(a_1, a_2, \cdots, a_m)=\frac{m!}{m_1!m_2! \cdots m_j!}.
$$

\begin{lem}  \label{yprimeovery}
$$
(y'/y)^{(n)}=\sum_{a_1 \geq a_2 \geq \cdots \geq a_m \geq 1:
a_1+a_2+\cdots+a_m=n+1} C_{[a_1, a_2, \cdots, a_m]} (y^{(a_1)}
y^{(a_2)} \cdots y^{(a_m)})/y^m,
$$
where $C_{[a_1, a_2, \cdots, a_m]}=(-1)^{m+1} \frac{1}{m} P(a_1,
a_2, \cdots, a_m) \frac{(a_1+a_2+\cdots+a_m)!}{a_1!a_2! \cdots
a_m!}$.
\end{lem}

\begin{proof}

One checks that $C_{[1]}=1$ and $C_{[a_1, a_2, \cdots, a_m]}$
satisfies the following recursion relationship:

For $a_1 \geq a_2 \geq \cdots \geq a_m \geq 2$,
\begin{equation}  \label{recursion_1}
C_{[a_1, a_2, \cdots, a_m]}=\sum D(a_1, a_2, \cdots, a_m;b_1, b_2,
\cdots, b_m) C_{[b_1, b_2, \cdots, b_m]},
\end{equation}
where the summation is over all $b_1 \geq b_2 \geq \cdots \geq b_m
\geq 1$, and all $b_i$ is equal to $a_i$ except for one of them, say
$b_k=a_k-1$, and $D(a_1, a_2, \cdots, a_m;b_1, b_2, \cdots, b_m)$ is
defined to the number of $b_k$ occurring in the sequence of $b_1,
b_2, \cdots, b_m$. For $a_1 \geq a_2 \geq \cdots \geq a_m=1$,
\begin{equation}  \label{recursion_2}
C_{[a_1, a_2, \cdots, a_m]}=\sum D(a_1, a_2, \cdots, a_m;b_1, b_2,
\cdots, b_m) C_{[b_1, b_2, \cdots, b_m]}-(m-1) C_{[a_1, a_2, \cdots,
a_{m-1}]};
\end{equation}
again here the summation is over all $b_1 \geq b_2 \geq \cdots \geq
b_m \geq 1$, and all $b_i$ is equal to $a_i$ except for one of them,
say $b_k=a_k-1$, and $D(a_1, a_2, \cdots, a_m;b_1, b_2, \cdots,
b_m)$ is defined to the number of $b_k$ occurring in the sequence of
$b_1, b_2, \cdots, b_m$.

One checks that
$$
(-1)^{m+1} \frac{1}{m} P(a_1, a_2, \cdots, a_m)
\frac{(a_1+a_2+\cdots+a_m)!}{a_1!a_2! \cdots a_m!},
$$
satisfies the initial value and recursion (\ref{recursion_1}) and
(\ref{recursion_2}). Since the initial value and recursion uniquely
determine the sequence, the theorem then follows.
\end{proof}

\begin{lem}  \label{highpart}
For $i=\lceil (N+1)/2 \rceil, \cdots, N$, $q_i[y]$ is proportional
to $(\log y +1)^{(N-i)}$. More specifically, we have
$$
q_i[y]=C_{i, N} (\log y +1)^{(N-i)},
$$
where $C_{i, N}$ is an integer.
\end{lem}

\begin{proof}

We first prove that for $N=2k+1$, the coefficient of $y^{(k+1)}$ is
proportional to $z^{(k-1)}$, where $z=(\log y +1)'=y'/y$. According
to Leibnitz formula, we have
$$
(y \log y)^{(2k+1)}=(y' (\log y+1))^{(2k)}=\sum_{l=0}^{2k} C_{2k}^l
y^{(l+1)} (\log y+1)^{(2k-l)}
$$
$$
=y^{(2k+1)} (\log y+1)+ \sum_{l=0}^{2k-1} C_{2k}^l y^{(l+1)}
z^{(2k-l-1)}.
$$
It suffices to prove that the coefficient of $y^{(k+1)}$ of
$$
C_{2k}^{k+1} y^{(k)} z^{(k)}+C_{2k}^{k+2} y^{(k-1)} z^{(k+1)}+
\cdots+ C_{2k}^{2k} y' z^{(2k-1)}
$$
is $C_{2k}^{k+1} z^{(k-1)}$. Applying Lemma~\ref{yprimeovery} and
collecting terms, we have the coefficient of $y^{(k+1)}$ equal to
$$
C_{2k}^{k+1} C_{[k+1]} y^{(k)}/y+C_{2k}^{k+2} C_{[k+1, 1]}
(y^{(k-1)} y^{(1)})/y^2
$$
$$
+C_{2k}^{k+3} C_{[k+1, 2]} (y^{(k-2)} y^{(2)})/y^2 + C_{2k}^{k+3}
C_{[k+1, 1, 1]} (y^{(k-2)} y^{(1)} y^{(1)})/y^3
$$
$$
+C_{2k}^{k+4} C_{[k+1, 3]} (y^{(k-3)} y^{(3)})/y^2+C_{2k}^{k+4}
C_{[k+1, 2, 1]} (y^{(k-3)} y^{(2)} y^{(1)})/y^3 + C_{2k}^{k+4}
C_{[k+1, 1, 1, 1]} (y^{(k-3)} y^{(1)} y^{(1)} y^{(1)})/y^4
$$
$$
+\cdots + C_{2k}^{2k} C_{[k+1, k-1]} (y^{(1)}
y^{(k-1)}/y^2+\cdots+C_{2k}^{2k} C_{[k+1, 1, \cdots, 1]} (y^{(1)}
y^{(1)} \cdots y^{(1)})/y^k.
$$

Consider the term $(y^{(a_1)} y^{(a_2}) \cdots y^{(a_m)})/y^m$ (here
$a_1+a_2+\cdots+a_m=k$) and compute its coefficient in the
expression above. Assuming that $a_1 \geq a_2 \geq \cdots \geq a_m$
satisfy (\ref{aaa}), we have the coefficient of $y^{(k+1)}$:
$$
C_{2k}^{2k+1-a_1} C_{[k+1, a_2, \cdots,
a_m]}+C_{2k}^{2k+1-a_{m_1+1}} C_{[k+1, a_1, \cdots, a_{m_1},
a_{m_1+2}, \cdots, a_m]}+\cdots
$$
$$
+C_{2k}^{2k+1-a_{m_1+m_2+\cdots+m_{j-1}+1}} C_{[k+1, a_1, \cdots,
a_{m_1+m_2+\cdots+m_{j-1}}, a_{m_1+m_2+\cdots+m_{j-1}+2}, \cdots,
a_m]}
$$
$$
=(-1)^{m+1} \frac{1}{m} \left( \frac{(2k)!}{(2k+1-a_1)!(a_1-1)!}
\frac{(2k+1-a_1)!}{(k+1)! a_2! \cdots a_m!} \frac{m!}{(m_1-1)! m_2!
\cdots m_j!}\right.
$$
$$
+\frac{(2k)!}{(2k+1-a_{m_1+1})!(a_{m_1+1}-1)!}
\frac{(2k+1-a_{m_1+1})!}{a_1! \cdots a_{m_1}! (k+1)! a_{m_1+2}!
\cdots a_m!} \frac{m!}{m_1! (m_2-1)! \cdots m_j!}+\cdots+
$$
$$
\left.\hspace{-2.5cm}+\frac{(2k)!}{(2k+1-a_{m_1+\cdots+m_{j-1}+1})!(a_{m_1+\cdots+m_{j-1}+1}-1)!}
\frac{(2k+1-a_{m_1+\cdots+m_{j-1}+1})!}{a_1! \cdots
a_{m_1+\cdots+m_{j-1}}! (k+1)! a_{m_1+\cdots+m_{j_1}+2}! \cdots
a_m!} \frac{m!}{m_1! m_2! \cdots (m_j-1)!} \right)
$$
$$
=(-1)^{m+1} \frac{1}{m} \frac{(2k)!}{(k+1)!} \frac{m!}{m_1! \cdots
m_j!} \frac{m_1a_1+m_2a_{m_1+1}+\cdots+m_j
a_{m_1+\cdots+m_{j-1}+1}}{a_1!a_2! \cdots a_m!}
$$
$$
=(-1)^{m+1} \frac{1}{m} \frac{(2k)!}{(k+1)!(k-1)!}
\frac{(a_1+a_2+\cdots+a_m)!}{a_1!a_2!\cdots a_m!}
\frac{m!}{m_1!m_2!\cdots m_j!}
$$
$$
=C_{2k}^{k+1} C_{[a_1, a_2, \cdots, a_m]}.
$$
It then follows that the coefficient of $y^{(k+1)}$ is equal to
$C_{2k}^{k+1} z^{(k-1)}$.

One can do similar computations to prove that for $N=2k, 2k+1$, this
lemma holds for other derivatives. An alternative approach is to use
induction. Using the fact that the coefficient of $y^{(k+1)}$ is
proportional to $z^{(k-1)}$ (established above), one can prove by
induction that for the $2k$-th order derivative of $y \log y$, the
coefficient of $y^{(l)}$ is proportional to $(\log y+1)^{(2k-l)}$
for $l$ with $k+1 \leq l \leq 2k$; and for $2k+1$-th order
derivative of $y \log y$, the coefficient of $y^{(l)}$ is
proportional to $(\log y+1)^{(2k+1-l)}$ for $l$ with $k+2 \leq l
\leq 2k+1$.
\end{proof}

\begin{lem}  \label{lowpart}
$$
\Low_N [ax]= \sum_{i=0}^{\lceil (N-1)/2 \rceil} r_i[a]
x^{(i)}+\sum_{i=0}^{\lceil (N-1)/2 \rceil} s_i[x] a^{(i)},
$$
where $r_i[a]$ is a function of $a$ and its derivatives (up to order
$\lceil (N-1)/2 \rceil$), and $s_i[x]$ is a function of $x$ and its
derivatives (up to order $\lceil (N-1)/2 \rceil$). Also,
$$
s_0[x]=\Low_N [x].
$$
\end{lem}

\begin{proof}
By Leibnitz formula, we have
$$
((ax) \log (ax))^{(N)}= \sum_{i=0}^N C_N^i (ax)^{(i)} (\log
(ax))^{(N-i)}
$$
$$
=\sum_{i=0}^N C_N^i \sum_{j=0}^i C_i^j a^{(j)} x^{(i-j)} (\log a +
\log x)^{(N-i)}.
$$
Thus there exist a function of $a$ and its derivatives $t_i[a]$, and
a function of $x$ and its derivatives $w_i[x]$ such that
$$
((ax) \log (ax))^{(N)}=\sum_{i=0}^N t_i[a] x^{(i)}+\sum_{i=0}^N
w_i[x] a^{(i)},
$$
with $w_0[x]= (x \log x)^{(N)}$.

By Lemma~\ref{highpart}, we have
$$
\High_N [ax]= \sum_{i=\lceil (N+1)/2 \rceil}^N q_i[ax]
(ax)^{(i)}=\sum_{i=\lceil (N+1)/2 \rceil}^N C_{i, N} (\log a + \log
x+1)^{(N-i)} (ax)^{(i)}
$$
Thus we conclude that there exist a function of $a$ and its
derivatives $u_i[a]$, and a function of $x$ and its derivatives
$v_i[x]$ such that
$$
\High_N [ax]=\sum_{i=\lceil (N+1)/2 \rceil}^N u_i[a] x^{(i)} +
\sum_{i=\lceil (N+1)/2 \rceil}^N v_i[x] a^{(i)},
$$
with $v_0[x]=\High_N [x]$. Since
$$
\Low_N [ax] = ((ax) \log (ax))^{(N)}- \High_N [ax],
$$
existence of $r_i[a]$ and $s_i[x]$ then follows, and they depend on
the derivatives only up to $\lceil (N-1)/2 \rceil$, and
$s_0[x]=\Low_N [x]$.
\end{proof}

\begin{thm}  \label{stabilizingtheorem2}
If at $\varepsilon=\hat{\varepsilon}$, for every $a \in A$,
$\Delta_a$ is a rank one matrix, and every column of $\Delta_a$ is
either a positive or a zero column, then
$$
\hspace{-5mm} \left.
\frac{\partial^{\alpha_1+\alpha_2+\cdots+\alpha_m}
H(Z)}{\partial_{\varepsilon_1}^{\alpha_1}
\partial_{\varepsilon_2}^{\alpha_2} \cdots
\partial_{\varepsilon_m}^{\alpha_m}} \right |_{\varepsilon=\hat{\varepsilon}}=\left. \frac{\partial^{\alpha_1+\alpha_2+\cdots+\alpha_m}
H_{\lceil(\alpha_1+\alpha_2+\cdots+\alpha_m+1)/2\rceil}(Z)}{\partial_{\varepsilon_1}^{\alpha_1}
\partial_{\varepsilon_2}^{\alpha_2} \cdots
\partial_{\varepsilon_m}^{\alpha_m}} \right |_{\varepsilon=\hat{\varepsilon}}
$$
\end{thm}

\begin{proof}

For simplicity we assume that $\Delta$ is only parameterized by only
one variable $\varepsilon$, and we drop $\varepsilon$ when the
implication is clear from the context. Recall that
$$
H_n(Z)=-\sum_{z_{-n}^0} p(z_{-n}^0) \log
p(z_0|z^{-1}_{-n})=-(\sum_{z_{-n}^0}p(z_{-n}^0) \log p(z_{-n}^0) -
\sum_{z_{-n}^{-1}} p(z_{n}^{-1}) \log p(z_{n}^{-1})).
$$

With slight abuse of notation (by replacing the formal derivative
with the derivative with respect to $\varepsilon$, we can define
$\High_N [p(z_{-n}^0)]=\High_N [p^{\varepsilon}(z_{-n}^0)]$.
Similarly for $\Low_N [p(z_{-n}^0)]$, etc.),
$$
(p(z_{-n}^0) \log p(z_{-n}^0))^{(N)}=\High_N [p(z_{-n}^0)]+ \Low_N
[p(z_{-n}^0)]
$$
$$
(p(z_{-n}^{-1}) \log p(z_{-n}^{-1}))^{(N)}=\High_N [p(z_{-n}^{-1})]+
\Low_N [p(z_{-n}^{-1})]
$$
Note that by Lemma~\ref{highpart}, we have
$$
\High_N [p(z_{-n}^0)]=\sum_{i=\lceil (N+1)/2 \rceil}^N C_{i, N}
(\log p(z_0|z_{-n}^{-1})+\log p(z_{-n}^{-1})+1)^{(N-i)}
p(z_{-n}^0)^{(i)},
$$
and
$$
\High_N [p(z_{-n}^{-1})]=\sum_{i=\lceil (N+1)/2 \rceil}^N C_{i, N}
(\log p(z_{-n}^{-1})+1)^{(N-i)} p(z_{-n}^{-1})^{(i)}.
$$
Thus
$$
\sum_{z_{-n}^0} \High_N [p(z_{-n}^0)]- \sum_{z_{-n}^{-1}} \High_N
[p(z_{-n}^{-1})]
$$
$$
=\sum_{z_{-n}^0} \sum_{i=\lceil (N+1)/2 \rceil}^N C_{i, N} (\log
p(z_0|z_{-n}^{-1})+\log p(z_{-n}^{-1})-\log p(z_{-n}^{-1}))^{(N-i)}
p(z_{-n}^0)^{(i)}
$$
$$
=\sum_{z_{-n}^0} \sum_{i=\lceil (N+1)/2 \rceil}^N C_{i, N} (\log
p(z_0|z_{-n}^{-1}))^{(N-i)} p(z_{-n}^0)^{(i)}
$$
$$
=\sum_{z_{-n}^0} \sum_{i=\lceil (N+1)/2 \rceil}^N C_{i, N} (\log
p(z_0|z_{-\lceil (N+1)/2 \rceil}^{-1}))^{(N-i)} p(z_{-\lceil (N+1)/2
\rceil}^0)^{(i)}
$$
So the higher derivative part stabilizes at $\lceil (N+1)/2 \rceil$,
namely for any $n \geq \lceil (N+1)/2 \rceil$, $\sum_{z_{-n}^0}
\High_N [p(z_{-n}^0)]- \sum_{z_{-n}^{-1}} \High_N [p(z_{-n}^{-1})]$
is equal to $\sum_{z_{-\lceil (N+1)/2 \rceil}^0} \High_N
[p(z_{-\lceil (N+1)/2 \rceil}^0)]- \sum_{z_{-\lceil (N+1)/2
\rceil}^{-1}} \High_N [p(z_{-\lceil (N+1)/2 \rceil}^{-1})]$. And by
Lemma~\ref{lowpart}, we have
$$
\Low_N [p(z_{-n}^0)]= \sum_{i=0}^{\lceil (N-1)/2 \rceil}
r_i[p(z_0|z_{-n}^{-1})] p(z_{-n}^{-1})^{(i)}+\sum_{i=0}^{\lceil
(N-1)/2 \rceil} s_i[p(z_{-n}^{-1}))] p(z_0|z_{-n}^{-1})^{(i)},
$$
with $s_0[p(z_{-n}^{-1}))]=\Low_N [p(z_{-n}^{-1})]$. Thus,
$$
\sum_{z_{-n}^0} \Low_N [p(z_{-n}^0)]- \sum_{z_{-n}^{-1}} \Low_N
[p(z_{-n}^{-1})]
$$
$$
=\sum_{z_{-n}^0} \sum_{i=0}^{\lceil (N-1)/2 \rceil}
r_i[p(z_0|z_{-n}^{-1})] p(z_{-n}^{-1})^{(i)}.
$$
$$
=\sum_{z_{-n}^0} \sum_{i=0}^{\lceil (N-1)/2 \rceil}
r_i[p(z_0|z_{-\lceil (N+1)/2 \rceil}^{-1})] p(z_{-\lceil (N+1)/2
\rceil}^{-1})^{(i)}.
$$
Consequently the lower derivative part stabilizes at $\lceil (N+1)/2
\rceil$ as well, namely for any $n \geq \lceil (N+1)/2 \rceil$,
$\sum_{z_{-n}^0} \Low_N [p(z_{-n}^0)]- \sum_{z_{-n}^{-1}} \Low_N
[p(z_{-n}^{-1})]$ is equal to $\sum_{z_{-\lceil (N+1)/2 \rceil}^0}
\Low_N [p(z_{-\lceil (N+1)/2 \rceil}^0)]- \sum_{z_{-\lceil (N+1)/2
\rceil}^{-1}} \Low_N [p(z_{-\lceil (N+1)/2 \rceil}^{-1})]$. The
theorem then follows.

\end{proof}

\begin{rem}
For an irreducible stationary Markov chain $Y$ with probability
transition matrix $\Delta$, let $Y^{-1}$ denote its reverse Markov
chain. It is well known that the probability transition matrix of
$Y^{-1}$ is $\diag(\pi_1^{-1}, \pi_2^{-1}, \cdots, \pi_B^{-1})
\Delta^t \diag(\pi_1, \pi_2, \cdots, \pi_B)$, where $\Delta^t$
denotes the transpose of $\Delta$ and $(\pi_1, \pi_2, \cdots,
\pi_B)$ is the stationary vector of $Y$. Therefore if $\Delta^t$ is
a Black Hole case, the derivatives of $H(Z^{-1})$ (here, $Z^{-1}$ is
the reverse hidden Markov chain defined by $Z^{-1}=\Phi(Y^{-1})$)
also stabilize. It then follows from $H(Z)=H(Z^{-1})$ that the
derivatives of $H(Z)$ also stabilize.
\end{rem}

\section{Binary Markov Chains Corrupted by Binary Symmetric Noise}
\label{particular}

In this section, we further study hidden Markov chains obtained by
binary Markov chains corrupted by binary symmetric noise with
crossover probability $\eps$ (described in Example 4.1
of~\cite{gm05}).  We take a concrete approach to study $H(Z)$, and
we will ``compute'' $H'(Z)$ in terms of Blackwell's measure.

Here the Markov chain is defined by a $2 \times 2$ stochastic matrix
$\Pi = [\pi_{ij}]$ (the reader should not confuse $\Pi$ with the $4
\times 4$ matrix $\Delta$:
$$
\left [ \begin{array}{cccc}
\pi_{00} (1-\varepsilon) & \pi_{00} \varepsilon & \pi_{01} (1-\varepsilon) & \pi_{01} \varepsilon\\
\pi_{00} (1-\varepsilon) & \pi_{00} \varepsilon & \pi_{01} (1-\varepsilon) & \pi_{01} \varepsilon\\
\pi_{10} (1-\varepsilon) & \pi_{10} \varepsilon & \pi_{11} (1-\varepsilon) & \pi_{11} \varepsilon\\
\pi_{10} (1-\varepsilon) & \pi_{10} \varepsilon & \pi_{11} (1-\varepsilon) & \pi_{11} \varepsilon\\
\end{array} \right ],
$$
which defines the hidden Markov chain via a deterministic function).

When $\det(\Pi)=0$, the rows of $\Pi$ are identical, and so $Y$ is
an i.i.d. random sequence with distribution $(\pi_{00}, \pi_{01})$.
Thus, $Z$ is an i.i.d. random sequence with distribution $(\pi, 1
-\pi)$ where $\pi = \pi_{00} (1 - \varepsilon) + \pi_{01}
\varepsilon$. So,
$$
H(Z)=-\pi \log \pi-(1-\pi) \log(1-\pi).
$$

From now through the end of Section~\ref{xxxV-B}, we {\bf assume:}
\begin{itemize}
\item  $\det(\Pi) > 0$ -- and --
\item all $\pi_{ij} > 0$ -- and --
\item $\varepsilon > 0$.
\end{itemize}
We remark that the condition $\det(\Pi) > 0$ is purely for
convenience. Results in this section will hold with the condition
$\det(\Pi) < 0$ through similar arguments, unless specified
otherwise.

The integral formula (\ref{blackwell_form}) expresses $H(Z)$ in
terms of the measure $Q$ on the 4-dimensional simplex; namely $Q$ is
the distribution of $p((y_0, e_0)|z_{-\infty}^0)$.  However, in the
case under consideration, $H(Z)$ can be expressed as an integral on
the real line~\cite{or03}, which we review as follows.

From the chain rule of probability theory,
$$
p(z_1^i, y_i)=p(z_1^{i-1}, z_i, y_{i-1}=0, y_i)+p(z_1^{i-1}, z_i,
y_{i-1}=1, y_i)
$$
$$
=p(z_i, y_i | z_1^{i-1}, y_{i-1}=0) p(z_1^{i-1}, y_{i-1}=0)+p(z_i,
y_i | z_1^{i-1}, y_{i-1}=1) p(z_1^{i-1}, y_{i-1}=1),
$$
and
$$
p(z_i, y_i | z_1^{i-1}, y_{i-1}=0)=p(z_1^i | z_1^{i-1}, y_i,
y_{i-1}=0) p(y_i | z_1^{i-1}, y_{i-1}=0)
$$
$$
=p(z_i | y_i) p(y_i | y_{i-1}=0)=p_E(e_i) p(y_i | y_{i-1}=0).
$$
Let $a_i=p(z_1^i, y_i=0)$ and $b_i=p(z_1^i, y_i=1)$. The pair
$(a_i,b_i)$ satisfies the following dynamical system:
$$
\left \{ \begin{array}{c} a_i=p_E(z_i) \pi_{00} a_{i-1} +
p_E(z_i) \pi_{10} b_{i-1}\\
b_i=p_E(\bar{z}_i) \pi_{01} a_{i-1} + p_E(\bar{z}_i) \pi_{11}
b_{i-1}.\\
\end{array}\right.
$$

Let $x_i=a_i/b_i$, we have a dynamical system with just one
variable:
$$
x_{i+1}=f_{z_{i+1}}(x_i),
$$
where
$$
f_{z}(x)=\frac{p_E(z)}{p_E(\bar{z})} \frac{\pi_{00}
x+\pi_{10}}{\pi_{01} x+\pi_{11}}, \qquad  z=0, 1
$$
starting with
$$
x_0=\pi_{10}/\pi_{01}.
$$

We are interested in the invariant distribution of $x_n$, which is
closely related to Blackwell's distribution of $p((y_0,
e_0)|z_{-\infty}^0)$. Now
\begin{eqnarray*}
p(y_i=0 | z_1^{i-1})&=&p(y_i=0, y_{i-1}=0 | z_1^{i-1})+p(y_i=0, y_{i-1}=1 | z_1^{i-1})\\
&=&\pi_{00}p(y_{i-1}=0|z_1^{i-1})+\pi_{10}p(y_{i-1}=1|z_1^{i-1})\\
&=&\pi_{00} \frac{a_{i-1}}{a_{i-1}+b_{i-1}}+\pi_{10}
\frac{b_{i-1}}{a_{i-1}+b_{i-1}}\\
&=&\pi_{00} \frac{x_{i-1}}{1+x_{i-1}}+\pi_{10}
\frac{1}{1+x_{i-1}}.\\
\end{eqnarray*}

Similarly we have
\begin{eqnarray*}
p(y_i=1 | z_1^{i-1})&=&p(y_i=1, y_{i-1}=0 | z_1^{i-1})+p(y_i=1, y_{i-1}=1 | z_1^{i-1})\\
&=&\pi_{01} \frac{x_{i-1}}{1+x_{i-1}}+\pi_{11} \frac{1}{1+x_{i-1}}.\\
\end{eqnarray*}

Further computation leads to
\begin{eqnarray*}
p(z_i=0|z_1^{i-1})&=&p(y_i=0,e_i=0|z_1^{i-1})+p(y_i=1,e_i=1|z_1^{i-1})\\
&=&p(e_i=0) p(y_i=0|z_1^{i-1})+p(e_i=1) p(y_i=1|z_1^{i-1})\\
&=&((1-\varepsilon) \pi_{00}+\varepsilon \pi_{01})
\frac{x_{i-1}}{1+x_{i-1}}+ ((1-\varepsilon) \pi_{10}+ \varepsilon
\pi_{11}) \frac{1}{1+x_{i-1}}\\
&=&r_0(x_{i-1}),\\
\end{eqnarray*}
where

\begin{equation} \label{r0}
r_0(x)= \frac{((1-\varepsilon) \pi_{00}+\varepsilon \pi_{01})
x+((1-\varepsilon) \pi_{10}+ \varepsilon \pi_{11})}{x+1}.
\end{equation}
Similarly we have
\begin{eqnarray*}
p(z_i=1|z_1^{i-1})&=&p(y_i=0,e_i=1|z_1^{i-1})+p(y_i=1,e_i=0|z_1^{i-1})\\
&=&p(e_i=1) p(y_i=0|z_1^{i-1})+p(e_i=0) p(y_i=1|z_1^{i-1})\\
&=&((\varepsilon \pi_{00}+(1-\varepsilon) \pi_{01})
\frac{x_{i-1}}{1+x_{i-1}}+ (\varepsilon \pi_{10}+ (1-\varepsilon) \pi_{11}) \frac{1}{1+x_{i-1}}\\
&=&r_1(x_{i-1}),\\
\end{eqnarray*}
where

\begin{equation}  \label{r1}
r_1(x)=\frac{(\varepsilon \pi_{00}+(1-\varepsilon) \pi_{01})
x+(\varepsilon \pi_{10}+ (1-\varepsilon) \pi_{11})}{x+1}.
\end{equation}

Now we write
$$
p(x_i \in E | x_{i-1})= \sum_{\{a|f_a(x_{i-1}) \in E\}} p(z_i=a |
x_{i-1}).
$$

Note that
\begin{eqnarray*}
p(z_i=0 | x_{i-1})&=&p(z_i=0 | z_1^{i-1})=r_0(x_{i-1}),\\
p(z_i=1 |x_{i-1})&=&p(z_i=1 | z_1^{i-1})=r_1(x_{i-1}).
\end{eqnarray*}

The analysis above leads to
$$
p(x_i \in E)= \int_{f_0^{-1}(E)} r_0(x_{i-1})
dp(x_{i-1})+\int_{f_1^{-1}(E)} r_1(x_{i-1}) dp(x_{i-1}).
$$

Abusing notation, we let $Q$ denote the limiting distribution of
$x_i$ (the limiting distribution exists due to the martingale
convergence theorem) and obtain:
\begin{equation}   \label{invariant_1}
Q(E)=\int_{f_0^{-1}(E)} r_0(x) dQ(x)+\int_{f_1^{-1}(E)} r_1(x)
dQ(x).
\end{equation}

We may now compute the entropy rate of $Z_i$ in terms of $Q$. Note
that
\begin{eqnarray*}
E(\log p(z_i| z_1^{i-1}))&=&E(p(z_i=0|z_1^{i-1}) \log
p(z_i=0|z_1^{i-1}))+p(z_i=1|z_1^{i-1}) \log p(z_i=1|z_1^{i-1}))\\
&=& E(r_0(x_{i-1}) \log r_0(x_{i-1})+r_1(x_{i-1}) \log
r_1(x_{i-1})).\\
\end{eqnarray*}

Thus (\ref{blackwell_form}) becomes
\begin{equation} \label{blackwell_conc}
H(Z)=-\int (r_0(x) \log r_0(x)+r_1(x) \log r_1(x)) dQ(x).
\end{equation}

\subsection{Properties of $Q$}

Since $\det(\Pi) > 0$, $f_0$ and $f_1$ are increasing continuous
functions bounded from above, and $f_0(0)$ and $f_1(0)$ are
positive; therefore they each have a unique positive fixed point,
$p_0$ and $p_1$. Since $f_1$ is dominated by $f_0$, we conclude $p_1
\leq p_0$. Let
\begin{itemize}
\item
$I$ denote the interval $[p_1, p_0]$ -- and --
\item
$L=\bigcup_{n=1}^\infty L_n$ where
$$
L_n=\{f_{i_1} \circ f_{i_2} \cdots \circ f_{i_n}(p_j) | i_1, i_2,
\cdots, i_n \in \{0, 1\}, j=0, 1\}.
$$
\end{itemize}

Let $I_{i_1 i_2 \cdots i_n}$ denote $f_{i_n} \circ f_{i_{n-1}} \circ
\cdots \circ f_{i_1}(I)$, and $p_{i_1 i_2 \cdots i_n}$ denote
$p(z_1=i_1, z_2=i_2, \cdots, z_n=i_n)$. The {\em support} of a
probability measure $Q$, denoted $supp(Q)$, is defined as the
smallest closed subset with measure one.

\begin{thm}  \label{Support}
$supp(Q)=\bar{L}$.
\end{thm}

\begin{proof}
First, by straightforward computation, one can check that
$f'_0(p_0)$ and $f'_1(p_1)$ are both less than $1$.  Thus, $p_0$ and
$p_1$ are attracting fixed points.  Since $p_i$ is the unique
positive fixed point of $f_i$, it follows that the entire positive
half of the real line is in the domain of attraction of each $f_i$,
i.e. for any $p
>0$,     $f_i^{(n)}(p)$ approaches $p_i$
(here the superscript ${ }^{(n)}$ denotes the composition of $n$
copies of the function).

We claim that both $p_0$ and $p_1$ are in $supp(Q)$. If $p_0$ is not
in the support, then there is a neighborhood $I_{p_0}$ containing
$p_0$ with $Q$-measure $0$.  For any point $p >0$, for some $n$,
$f_0^{(n)}(p) \in I_{p_0}$.  Thus, by Equation~\ref{invariant_1}
there is a neighborhood of $p$ with $Q$-measure $0$. It follows that
$Q([0,\infty)) =0$.  On the other hand, $Q$ is the limiting
distribution of $x_i > 0$ and so   $Q([0,\infty)) =1$. This
contradiction shows that $p_0 \in supp(Q)$.  Similarly, $p_1 \in
supp(Q)$.

By Equation~\ref{invariant_1}, we deduce
$$
f_i(supp(Q)) \subseteq supp(Q).
$$
It follows that $L \subseteq supp(Q)$. Thus $\bar{L} \subseteq
supp(Q)$.

Since $f_i((0, \infty))$ is contained in a compact set, we may
assume $f_i$ is a contraction mapping (otherwise compose $f_0$ or
$f_1$ enough many times to make the composite mapping a contraction
as we argued in~\cite{gm05}). In this case the set of accumulation
points of $\{f_{i_n} \circ f_{i_{n-1}} \cdots \circ f_{i_1}(p) |
i_1, i_2, \cdots, i_n \in \{0, 1\}, p > 0\}$ does not depend on $p$.
Since any point in $supp(Q)$ has to be an accumulation point of
$\{f_{i_n} \circ f_{i_{n-1}} \cdots \circ f_{i_1}(\pi_{10}/\pi_{01})
| i_1, i_2, \cdots, i_n \in \{0, 1\}\}$, it has to be an
accumulation point of $L$ as well, which implies $supp(Q) \subseteq
\bar{L}$.
\end{proof}

It is easy to see that:

\begin{lem}
\label{equiv} The following statements are equivalent.
\begin{enumerate}
\item $f_0(I) \cup f_1(I) \subsetneqq I$.
\item $f_0(I) \cap f_1(I) = \phi$.
\item $f_1(p_0) < f_0(p_1)$.
\end{enumerate}
\end{lem}

\begin{thm} \label{Cantor-Interval}
$supp(Q)$ is either a Cantor set or a closed interval. Specifically:
\begin{enumerate}
\item $supp(Q)$ is a Cantor set if $f_0(I) \cup f_1(I) \subsetneqq I$.
\item $supp(Q)=I$ if equivalently $f_0(I) \cup f_1(I)=I$.
\end{enumerate}
\end{thm}

\begin{proof}
Suppose that $f_0(I) \cup f_1(I) \subsetneqq I$. If $(i_1, i_2,
\cdots, i_n) \neq (j_1, j_2, \cdots, j_n)$, then
$$
I_{i_1 i_2 \cdots i_n} \cap I_{j_1 j_2 \cdots j_n}=\phi.
$$
Define:
$$
I_{<n>}= \bigcup_{i_1, i_2, \cdots, i_n} I_{i_1 i_2 \cdots i_n}.
$$
Alternatively we can construct $I_{<n>}$ as follows: let
$I^d=(f_1(p_0), f_0(p_1) )$, then
$$
I_{<n+1>}=I_{<n>} \backslash \bigcup_{i_1, i_2, \cdots, i_n} f_{i_n}
\circ f_{i_{n-1}} \circ \cdots \circ f_{i_1} (I^d).
$$
Let $I_{<\infty>}= \bigcap_{n=1}^{\infty} I_{<n>}$. It follows from
the way it is constructed that $I_{\infty}$ is a Cantor set (think
of $I^d$ as a ``deleted'' interval), and $\bar{L}=I_{<\infty>}$.
Thus by Theorem~\ref{Support} $supp(Q)=\bar{L}$ is a Cantor set.

Suppose $f_0(I) \cup f_1(I) = I$. In this case, for any point $p \in
I$, and for all $n$, there exists $i_1, i_2, \cdots, i_n$ such that
$$
p \in I_{i_1 i_2 \cdots i_n}.
$$
From the fact that $f_0$ and $f_1$ are both contraction mappings
(again, otherwise compose $f_0$ or $f_1$ enough many times to make
the composite mapping a contraction as we argued in~\cite{gm05}), we
deduce that the length of $I_{i_1 i_2 \cdots i_n}$ is exponentially
decreasing with respect to $n$. It follows that $L$ is dense in $I$,
and therefore $supp(Q)=\bar{L}=I$.
\end{proof}

\begin{thm}  \label{continuity}
$Q$ is a continuous measure, namely for any point $p \in supp(Q)$,
and for any $\eta > 0$, there exists an interval $I_p$ containing
$p$ with $Q(I_p) < \eta$ (or equivalently $Q$ has no point mass).
\end{thm}

\begin{proof}
Assume that there exists $p \in I$ such that for any interval
containing $p$, $Q(I_p) > \eta_0$, where $\eta_0$ is a positive
constant. Let $\xi=\max\{r_0(x), r_1(x): x \in I\}$. One checks that
$0 < \xi < 1$. By (\ref{invariant_1}), we have
$$
\frac{1}{\xi} Q(I_p) \leq Q(f_0^{-1}(I_p))+ Q(f_1^{-1}(I_p)).
$$
Iterating, we obtain
$$
\left(\frac{1}{\xi}\right)^n \eta_0 \leq \sum_{i_1, i_2, \cdots,
i_n} Q( f_{i_1}^{-1} \circ f_{i_2}^{-1} \circ \cdots \circ
f_{i_n}^{-1}(I_p)).
$$
For fixed $n$, if we choose $I_p$ small enough, then
$$
f_{i_1}^{-1} \circ f_{i_2}^{-1} \circ \cdots \circ f_{i_n}^{-1}(I_p)
\cap f_{j_1}^{-1} \circ f_{j_2}^{-1} \circ \cdots \circ
f_{j_n}^{-1}(I_p)=\phi,
$$
for $(i_1, i_2, \cdots, i_n) \neq (j_1, j_2, \cdots, j_n)$. It
follows in this case that
$$
Q(I) \geq \sum_{i_1, i_2, \cdots, i_n} Q(( f_{i_1}^{-1} \circ
f_{i_2}^{-1} \circ \cdots \circ f_{i_n}^{-1}(I_p)) \geq
\left(\frac{1}{\xi}\right)^n \eta_0.
$$
Therefore for large $n$, we deduce
$$
Q(I) > 1,
$$
which contradicts the fact that $Q$ is a probability measure.
\end{proof}

By virtue of   Lemma~\ref{equiv}, it makes sense to refer to case
$1$ in Theorem~\ref{Cantor-Interval} as the {\em non-overlapping}
case. We now focus on this case. Note that this is the case whenever
$\varepsilon$ is sufficiently small; also, it turns out that for
some values of $\pi_{ij}$'s, the non-overlapping case holds for all
$\varepsilon$.

Starting with $x_0 = \pi_{10}/\pi_{01}$, and iterating according to
$x_n=f_{z_n}(\varepsilon, x_{n-1})$, each word $z = z_1, z_2,
\cdots, z_n$ determines a point $x_n = x_n(z)$ with probability
$p(z_1, z_2, \cdots, z_n)$. In the non-overlapping case, the map $z
\mapsto x_n$ is one-to-one.  We order the distinct points $\{x_n\}$
from left to right as
$$
x_{n, 1}, x_{n, 2}, \cdots, x_{n, 2^n}
$$
with the associated probabilities
$$
p_{n, 1}, p_{n, 2}, \cdots, p_{n, 2^n}.
$$
This defines a sequence of distribution $Q_n$ which converge weakly
to $Q$. In particular, by the continuity of $Q$, $Q_n(J) \rightarrow
Q(J)$ for any interval $J$.

\begin{thm} \label{xxx}
In the non-overlapping case,
$$
Q(I_{i_1 i_2 \cdots i_n})=Q_n(I_{i_1 i_2 \cdots i_n})=p_{i_1 i_2
\cdots i_n}.
$$
\end{thm}

\begin{proof}
We have
$$
Q_n(I_{i_1 i_2 \cdots i_n})=p(z_1=i_1, z_2=i_2, \cdots, z_n=i_n).
$$
Furthermore
$$
Q_{n+1}(I_{i_1 i_2 \cdots i_n})= Q_{n+1}(I_{0 i_1 i_2 \cdots
i_n})+Q_{n+1}(I_{1 i_1 i_2 \cdots i_n})
$$
$$
=p(z_0=0, z_1=i_1, z_2=i_2, \cdots, z_n=i_n)+p(z_0=1, z_1=i_1,
z_2=i_2, \cdots, z_n=i_n)
$$
$$
=p(z_1=i_1, z_2=i_2, \cdots, z_n=i_n)
$$
Iterating one shows that for $m \geq n$,
$$
Q_m(I_{i_1 i_2 \cdots i_n})=Q_n(I_{i_1 i_2 \cdots i_n})=p_{i_1 i_2
\cdots i_n}.
$$
By the continuity of $Q$ (Theorem~\ref{continuity})
$$
Q(I_{i_1 i_2 \cdots i_n})=p_{i_1 i_2 \cdots i_n}.
$$
\end{proof}

From this, as in~\cite{or03, or04} we can derive bounds for the
entropy rate. Let $$r(x)=-(r_0(x) \log r_0(x)+r_1(x) \log r_1(x)).$$
Using (\ref{blackwell_conc}) and Theorem~\ref{xxx}, we obtain:
\begin{thm}
In the non-overlapping case,
$$
\sum_{i_1 i_2 \cdots i_n} r_{i_1 i_2 \cdots i_n}^m p_{i_1 i_2 \cdots
i_n} \leq H(Z) \leq \sum_{i_1 i_2 \cdots i_n} r_{i_1 i_2 \cdots
i_n}^M p_{i_1 i_2 \cdots i_n},
$$
where $r_{i_1 i_2 \cdots i_n}^m=\min_{x \in I_{i_1 i_2 \cdots i_n}}
r(x)$ and $r_{i_1 i_2 \cdots i_n}^M=\max_{x \in I_{i_1 i_2 \cdots
i_n}} r(x)$.
\end{thm}

\begin{proof}
This follows immediately from the formula for the entropy rate
$H(Z)$ (~\ref{blackwell_conc}).
\end{proof}

\subsection{Computation of the first derivative in non-overlapping case}  \label{xxxV-B}

To emphasize the dependence on $\varepsilon$, we write
$p_{n,i}(\varepsilon) = p_{n,i}$, $x_{n,i}(\varepsilon) = x_{n,i}$,
 $p_0(\varepsilon)  = p_0$, $p_1(\varepsilon)  = p_1$, and $Q_n(\varepsilon) = Q_n$.
Let $F_n(\varepsilon, x)$ denote the cumulative distribution
function of $Q_n(\varepsilon)$. Let $H_n^\varepsilon(Z)$ be the
finite approximation to $H^\varepsilon(Z)$. It can be easily checked
that
$$
H_n^\varepsilon(Z)=\int_I r(\varepsilon, x) dQ_n(\varepsilon)
$$
and we can rewrite (\ref{blackwell_conc}) as
$$
H^\varepsilon(Z)=\int_I r(\varepsilon, x) dQ(\varepsilon).
$$
In Theorem~\ref{hpz}, we express the derivative of the entropy rate,
with respect to $\varepsilon$, as the sum of four terms which have
meaningful interpretations. Essentially we are differentiating
$H^\varepsilon(Z)$ with respect to $\varepsilon$ under the integral
sign, but care must be taken since $Q(\varepsilon)$ is generally
singular and varies with $\varepsilon$.

Rewriting this using the Riemann-Stieltjes integral and applying
integration by parts, we obtain
\begin{eqnarray*}
H_n^\varepsilon(Z)&=&\int_I r(\varepsilon, x) dF_n(\varepsilon, x)\\
&=&F_n(\varepsilon, x) r(\varepsilon, x)|_{p_1(\varepsilon)}^{p_0(\varepsilon)}-\int_I F_n(\varepsilon, x) g(\varepsilon, x) dx,\\
\end{eqnarray*}
where $g(\varepsilon, x)=\frac{\partial r(\varepsilon, x)}{\partial
x}$.

{\it From now on ${}'$ denotes the derivative with respect to
$\varepsilon$}.
 Now,
$$
H_n^\varepsilon(Z)' =  r(\varepsilon, p_0(\varepsilon))' -
D_n(\varepsilon),
$$
where
$$
\hspace{ -1 cm} D_n(\varepsilon) =  \lim_{h \to 0} \frac{\int_I
F_n(\varepsilon+h, x) g(\varepsilon+h, x) dx-\int_I F_n(\varepsilon,
x) g(\varepsilon, x) dx}{h}.
$$
We can decompose $D_n(\varepsilon)$ into two terms:
$$
D_n(\varepsilon) = D_n^1(\varepsilon) + D_n^2(\varepsilon),
$$
where
$$
D_n^1(\varepsilon) = \lim_{h \to 0} \int_I \frac{F_n(\varepsilon+h,
x)-F_n(\varepsilon, x)}{h}g(\varepsilon, x)dx,
$$
and
$$
D_n^2(\varepsilon) = \int_I F_n(\varepsilon, x) g'(\varepsilon, x)
dx.
$$
In order to compute $D_n^1(\varepsilon)$, we partition $I$ into two
pieces: 1) small  intervals $(x_{n, i}(\varepsilon), x_{n,
i}(\varepsilon+h))$  and 2) the complement of the union of these
neighborhoods, to yield:
$$D_n^1(\varepsilon) = \lim_{h \to 0} \int_I \frac{F_n(\varepsilon+h, x)-F_n(\varepsilon,
x)}{h}g(\varepsilon, x)dx =$$
$$
-\sum_i p_{n, i}(\varepsilon) x_{n, i}(\varepsilon)' g(\varepsilon,
x_{n, i})(\varepsilon) +\int_I F'_n(\varepsilon, x) g(\varepsilon,
x) dx.
$$
Combining the foregoing expressions, we arrive at an expression for
 $H^\varepsilon_n(Z)'$:
$$
H^\varepsilon_n(Z)'=r(\varepsilon, p_0(\varepsilon))'
    +\sum_i p_{n, i}(\varepsilon) x'_{n, i}(\varepsilon) g(\varepsilon, x_{n,i}(\varepsilon))
   $$
   $$
   - \int_I
F'_n(\varepsilon, x) g(\varepsilon, x) dx -
 \int_I F_n(\varepsilon, x) g'(\varepsilon, x) dx.
$$

Write $H^\varepsilon(Z) = H(Z)$, $Q(\varepsilon) =Q$ and let
$F(\varepsilon,x)$ be the cumulative distribution function of
 $Q(\varepsilon)$.

We then show that $H^\varepsilon_n(Z)$ converges uniformly to
$H^\varepsilon(Z)$ and $H^\varepsilon_n(Z)'$ converges uniformly to
some function; it follows that this function is $H^\varepsilon(Z)'$.
This requires showing that the integrands in the second and third
terms of the previous expression converge to well-defined functions.

We think of the $x_{n,i}(\varepsilon)$ as {\em locations} of point
masses.  So, we can think of $x_{n,i}(\varepsilon)'$ as an
instantaneous location change.

\begin{enumerate}
\item \textbf{2nd term, Instantaneous Location Change (See Appendix~\ref{K-1}):}
For $x \in supp(Q(\varepsilon))$ and any sequence of points $x_{n_1,
i_1}(\varepsilon), x_{n_2, i_2}(\varepsilon), \cdots$ approaching
$x$, $K_1(\varepsilon, x)=\lim_{j \to \infty} x'_{n_j,
i_j}(\varepsilon)$ is a well-defined continuous function.
\item \textbf{3rd term, Instantaneous Probability Change (See Appendix~\ref{K-2}):}
Recall that $supp(Q(\varepsilon))$ is a Cantor set defined by a
collection of ``deleted'' intervals: namely, $I^d \equiv (f_0(p_1),
f_1(p_0))$, and all intervals of the form $f_{i_1} \circ f_{i_2}
\circ \cdots \circ f_{i_n} (I^d)$ (called deleted intervals on level
$n$). For $x$ belonging to a deleted interval on level $n$, define
$K_2(\varepsilon, x)=F'_n(\varepsilon, x)$. Since the union of
deleted intervals is dense in $I$, we can extend  $K_2(\varepsilon,
x)$ to a function on all $x \in I$, and we show that
$K_2(\varepsilon, x)$ is a well-defined continuous function.
\end{enumerate}
Using the boundedness of the instantaneous location change and
probability change (established in Appendix~\ref{Location-Change}
and Appendix~\ref{Probability-Change}) and the Arzela-Ascoli Theorem
(note that Appendix~\ref{K-1} and Appendix~\ref{K-2} imply pointwise
convergence of $H_n^{\varepsilon}(Z)'$ and
Appendix~\ref{Location-Change}, and
Appendix~\ref{Probability-Change} imply equicontinuity of
$H_n^{\varepsilon}(Z)'$), we obtain uniform convergence of
$H^\varepsilon_n(Z)'$ to $H^\varepsilon(Z)'$, which gives the
result:
\begin{thm}
\label{hpz}
 In the non-overlapping case,
$$
H^\varepsilon(Z)'=r(\varepsilon, p_0(\varepsilon))'
+\int_{supp(Q(\varepsilon))} K_1(\varepsilon, x) g(\varepsilon, x)
dF(\varepsilon,x)
$$
$$
-\int_I K_2(\varepsilon, x) g(\varepsilon, x)dx - \int_I
F(\varepsilon, x) g'(\varepsilon, x) dx.
$$
\end{thm}
Note that the second term in this expression is a weighted mean of
the instantaneous location change and   the third term in this
expression is a weighted mean of the instantaneous probability
change.

\begin{rem}
Using the same technique, we can give a similar formula for the
derivative of $H^\varepsilon(Z)$ with respect to $\pi_{ij}$'s when
$\varepsilon
> 0$. We can also give such formulae for higher derivatives in a similar way.
\end{rem}

\begin{rem}
The techniques in this section can be applied to give an expression
for the derivative of the entropy rate in the special overlapping
case where $f_0(p_1)=f_1(p_0)$.
\end{rem}

\subsection{Derivatives in other cases}  \label{xxxV-C}

\textbf{1.} If \textbf{any two of the} $\mathbf{\pi_{ij}}$'s
\textbf{are equal to} $\mathbf{0}$, then
$$
H^\varepsilon(Z)=-\varepsilon \log \varepsilon- (1-\varepsilon) \log
(1-\varepsilon)
$$
$H^\varepsilon(Z)$ is not differentiable with respect to
$\varepsilon$ at $\varepsilon=0$.

\textbf{2.} Of more interest, it was shown in~\cite{or04} that
$H(Z)$ is not differentiable with respect to $\varepsilon$ at
$\varepsilon=0$ when \textbf{exactly one of the}
$\mathbf{\pi_{ij}}$'s \textbf{is equal to} $\mathbf{0}$. We briefly
indicate how this is related to (\ref{blackwell_conc}). Consider the
case: $\pi_{00}=0$, $\pi_{01}=1$, $0 < \pi_{10} < 1$. Then for
$\varepsilon > 0$,
$$
H(Z) = - \int_{I_0} r_0(x) \log r_0(x) dQ -\int_{I_1} r_0(x) \log
r_0(x) dQ
$$
$$
- \int_{I_0} r_1(x) \log r_1(x) dQ -\int_{I_1} r_1(x) \log r_1(x)
dQ.
$$
When $\varepsilon \to 0$, the lengths of $I_0$ and $I_1$ shrink to
zero with $I_1$ approaching $0$ and $I_0$ approaching $\infty$. So,
of the four terms above, as $\varepsilon \to 0$, the dominating term
will be
$$
\int_{I_0} r_0(x) \log r_0(x) dQ \sim \varepsilon \log \varepsilon,
$$
and all the other three terms are bounded by $O(\varepsilon)$ (see
(\ref{r0}) and (\ref{r1})).  This indicates that $H(Z)$ is not
differentiable with respect to $\varepsilon$ at $\varepsilon=0$.

\textbf{3.} Consider the case that $\mathbf{\varepsilon = 0}$ and
all the $\pi_{ij}$'s {\bf are positive}. As discussed in Example 4.1
of~\cite{gm05}, the entropy rate is analytic as a function of
$\varepsilon$ and $\pi_{ij}$'s.

In~\cite{ja04} (and more generally in~\cite{zu04},~\cite{zu05}), an
explicit formula was given for $H'(Z)$ at $\mathbf{\varepsilon = 0}$
in this case. We briefly indicate how this is related to our results
in Section~\ref{xxxV-B}.

Instead of considering the dynamics of $x_n$ on the real line, we
consider those of $(a_n, b_n)$ on the $1$ dimensional simplex
$$
W=\{(w_1, w_2) : w_1+w_2=1, w_i \geq 0\}.
$$
Let $Q$ denote the limiting distribution of $(a_n, b_n)$ on $W$, the
entropy $H(Z)$ can be computed as follows
$$
H(Z)=\int_W -(r_0(w) \log r_0(w)+r_1(w) \log r_1(w)) dQ,
$$
where
$$
r_0(w)=((1-\varepsilon) \pi_{00}+\varepsilon \pi_{01})
w_1+((1-\varepsilon) \pi_{10}+ \varepsilon \pi_{11})w_2,
$$
$$
r_1(w)=((\varepsilon \pi_{00}+(1-\varepsilon) \pi_{01})
w_1+(\varepsilon \pi_{10}+ (1-\varepsilon) \pi_{11})w_2.
$$
In order to calculate the derivative, we split the  region of
integration into two disjoint parts $W=W^0 \cup W^1$ with
$$
W^0=\{t(0,1)+(1-t)(1/2, 1/2) : 0 \leq t \leq 1\},
$$
$$
W^1=\{t(1/2,1/2)+(1-t)(1, 0) : 0 \leq t \leq 1\}.
$$
Let $ r(w)=-(r_0(w) \log r_0(w)+r_1(w) \log r_1(w))$, and
$H^i(Z)=\int_{W^i} r(w) dQ$, then
$$
H(Z)=H^0(Z)+H^1(Z).
$$

For $W^0$, we represent every point $(w_1, w_2)$ using the
coordinate $w_1/w_2$. For $W^1$, we represent every point $(w_1,
w_2)$ using the coordinate $w_2/w_1$. Computation shows that
$H_n^{\varepsilon}(Z)$ uniformly converge to $H^{\varepsilon}(Z)$ on
$[0, 1/2]$. Note that expressions in Theorem~\ref{hpz} are not
computable for $\varepsilon > 0$, however we can apply similar
uniform convergence ideas in each of these regions to recover the
formula given in~\cite{ja04} for $\varepsilon=0$.

\textbf{4.} ({\bf Low SNR regime,} $\mathbf{ \varepsilon = 1/2}$) In
Corollary 6 of ~\cite{or03}, it was shown that in the symmetric case
(i.e., $\pi_{01} = \pi_{10}$), the entropy rate approaches zero at
rate $(1/2 - \varepsilon)^4$ as $\varepsilon$ approaches  1/2. It
can be shown that the entropy rates at $\varepsilon$ and $1 -
\varepsilon$ are the same, and so all odd order derivatives vanish
at $\varepsilon = 1/2$. It follows that this result of~\cite{or03}
is equivalent to the statement that in the symmetric case
$H''(Z)|_{\varepsilon=1/2} = 0$. We generalize this result to the
non-symmetric case as follows:
$$
H''(Z)|_{\varepsilon=1/2}=-4
\left(\frac{\pi_{10}-\pi_{01}}{\pi_{10}+\pi_{01}} \right)^2.
$$
For more details, see Appendix~\ref{Low-SNR}.

\section*{Appendices}\appendix

\section{Proof of Boundedness of Instantaneous Location Change}   \label{Location-Change}

\textbf{Claim:} For any fix $0 < \eta < 1/2$, $x^{(k)}_{n,
i}(\varepsilon) \leq C_1(k, \eta)$, $\eta \leq \varepsilon \leq
1/2$, $C_1$ is a positive constant only depending on $k, \eta$.

\begin{proof}
We only prove the case when $k=1$. Consider the iteration,
$$
x_{n+1}=f_{z_{n+1}}(\varepsilon, x_n).
$$
Take the derivative with respect to $\varepsilon$, we obtain
$$
x_{n+1}'=\frac{\partial f_{z_{n+1}}}{\partial
\varepsilon}(\varepsilon, x_n)+\frac{\partial f_{z_{n+1}}}{\partial
x}(\varepsilon, x_n) x_n'.
$$

Note that $\frac{\partial f_{z_{n+1}}}{\partial
\varepsilon}(\varepsilon, x_n)$ is uniformly bounded by a constant
and $\frac{\partial f_{z_{n+1}}}{\partial x}(\varepsilon, x_n)$ is
bounded by $\rho$ with $0 < \rho < 1$, we conclude $x_n'$ is
uniformly bounded too.
\end{proof}

\section{Proof of Boundedness of Instantaneous Probability Change}  \label{Probability-Change}

\textbf{Claim:} For $x \notin \{x_{n, i}\}$ and $0 \leq \varepsilon
\leq 1/2$, $F^{(k)}_n(\varepsilon, x) \leq C_2(k)$, where $C_2$ is a
positive constant only depending on $k$.

\begin{proof}
We only prove the case when $k=1$. For $x$ with $x_{n, 2i} < x <
x_{n, 2i+1}$, we have $F_n(\varepsilon, x)=F_{n-1}(\varepsilon, x)$,
and consequently $\frac{\partial F_n(\varepsilon, x)}{\partial
\varepsilon}=\frac{\partial F_{n-1}(\varepsilon, x)}{\partial
\varepsilon}$. For $x$ with $x_{n, 2i-1} < x < x_{n, 2i}$,
$\frac{\partial F_n(\varepsilon, x)}{\partial
\varepsilon}-\frac{\partial F_{n-1}(\varepsilon, x)}{\partial
\varepsilon}$ is bounded by $C\rho_1^n$, here $C$ is a positive
constant and $0 < \rho_1 < 1$ (see proof that $K_2$ is well-defined
in Appendix~\ref{K-2}). Therefore we conclude the instantaneous
probability change is uniformly bounded.
\end{proof}

\section{Proof that $K_1$ is Well-defined}  \label{K-1}
\begin{proof}
We need to prove that if two points $x_{n_k, i_k}$ and $x_{n_l,
i_l}$ are close, then $x'_{n_k, i_k}$ and $x'_{n_l, i_l}$ are also
close. Note that for non-overlapping case, if $x_{n_k, i_k}$ and
$x_{n_l, i_l}$ are very close, their corresponding symbolic
sequences must share a long common tail. We shall prove that the
asymptotical dynamics of $x_n$ does not depend on the starting point
as long as they have the same common long tail. Without loss of
generality, we assume that $z$, $\hat{z}$ have common tail $z_1,
z_2, \cdots, z_n$. In this case, the two dynamical systems start
with different value $x_0$, $\hat{x}_0$ along the same path. Now the
two iterations produce
$$
x_{n+1}'=\frac{\partial f_{z_{n+1}}}{\partial
\varepsilon}(\varepsilon, x_n)+\frac{\partial f_{z_{n+1}}}{\partial
x}(\varepsilon, x_n) x_n'.
$$
$$
\hat{x}_{n+1}'=\frac{\partial f_{z_{n+1}}}{\partial
\varepsilon}(\varepsilon, \hat{x}_n)+\frac{\partial
f_{z_{n+1}}}{\partial x}(\varepsilon, \hat{x}_n) \hat{x}_n'.
$$
Take the difference, we have
$$
x_{n+1}'-\hat{x}_{n+1}'=\frac{\partial f_{z_{n+1}}}{\partial
\varepsilon}(\varepsilon, x_n)- \frac{\partial f_{z_{n+1}}}{\partial
\varepsilon}(\varepsilon, \hat{x}_n)+ \frac{\partial
f_{z_{n+1}}}{\partial x}(\varepsilon, x_n) x_n'- \frac{\partial
f_{z_{n+1}}}{\partial x}(\varepsilon, \hat{x}_n) \hat{x}_n'
$$
$$
=\frac{\partial f_{z_{n+1}}}{\partial \varepsilon}(\varepsilon,
x_n)- \frac{\partial f_{z_{n+1}}}{\partial \varepsilon}(\varepsilon,
\hat{x}_n)+ \frac{\partial f_{z_{n+1}}}{\partial x}(\varepsilon,
x_n) x_n'- \frac{\partial f_{z_{n+1}}}{\partial x}(\varepsilon,
\hat{x}_n) x_n'+ \frac{\partial f_{z_{n+1}}}{\partial
x}(\varepsilon, \hat{x}_n) x_n' - \frac{\partial
f_{z_{n+1}}}{\partial x}(\varepsilon, \hat{x}_n) \hat{x}_n'
$$

Since
\begin{itemize}
\item when $n \to \infty$, $x_n$ and $\hat{x}_n$ are getting close
uniformly with respect to $\varepsilon$  -- and --
\item $\frac{\partial f_i}{\partial \varepsilon}(\varepsilon, \cdot)$ and $\frac{\partial
f_i}{\partial x}(\varepsilon, \cdot)$ ($i=0,1$) are Lipschitz  --
and --
\item  $f_i (\varepsilon, \cdot)$ ($i=0,1$) are $\rho$-contraction mappings,
\end{itemize}
we conclude that $x_n'$ and $\hat{x}_n'$ are very close uniformly
with respect to $\varepsilon$. The well-definedness of $K_1$ then
follows.
\end{proof}

\section{Proof that $K_2$ is Well-defined}  \label{K-2}
\begin{proof}

Every deleted interval corresponds to a finite sequence of binary
digits and $K_2$ is well defined on these intervals. We order the
deleted intervals on level $n$ from left to right
$$
I_{n, 1}^d, I_{n, 2}^d, \cdots, I_{n, 2^{n-1}}^d.
$$
We need to prove if two deleted intervals $I_{m, i}^d$, $I_{n, j}^d$
are close, then $F_m(\varepsilon, I_{m, i}^d)$ (which is defined as
$F_m(\varepsilon, x)$ with $x \in I_{m, i}^d$) and $F_m(\varepsilon,
I_{m, i}^d)$ are close. Assume $m \leq n$, then the points $x_{n,
k}$'s in between $I_{m, i}^d$ and $I_{n, j}^d$ must have a long
common tail. Suppose that the common tail is the path $z_1, z_2,
\cdots, z_n$, let $q_i$ denote the sum of the probabilities
associated with these points. Note that as long as the sequences
have long common tail, the corresponding values of $K_2$ are getting
closer and closer. For simplicity we only track one path for the
time being. Then we have
\begin{eqnarray*}
a_{i+1}&=&p_E(z_{i+1})(\pi_{00} a_i+ \pi_{10} b_i),\\
b_{i+1}&=&p_E(\bar{z}_{i+1})(\pi_{01} a_i+ \pi_{11} b_i).\\
\end{eqnarray*}
It follows that
$$
(a_{i+1}+b_{i+1}) \leq \rho (a_i+b_i),
$$
here $0 < \rho < 1$ and $\rho$ is defined as
$$
\rho=\max \{ (1-\varepsilon) \pi_{00}+\varepsilon
\pi_{01},(1-\varepsilon) \pi_{10}+\varepsilon \pi_{11}, \varepsilon
\pi_{00}+(1-\varepsilon) \pi_{01}, \varepsilon
\pi_{10}+(1-\varepsilon) \pi_{11} \}.
$$
Immediately we have
$$
(a_n+b_n) \leq \rho^n.
$$
Take the derivative, we have
\begin{eqnarray*}
a_{n+1}'&=&-(\pi_{00} a_n + \pi_{10} b_n)+(1-\varepsilon)
(\pi_{00} a_n'+\pi_{10} b_n'),\\
b_{n+1}'&=&(\pi_{01} a_n + \pi_{11} b_n)+\varepsilon
(\pi_{10} a_n'+\pi_{11} b_n').\\
\end{eqnarray*}
In this case we obtain,
$$
|a_{n+1}'|+|b_{n+1}'| \leq \rho (|a_n'|+|b_n'|)+\rho^n,
$$
which implies that there is a positive constance $C$ and $\rho_1$
with $\rho < \rho_1 < 1$ such that
$$
a_n'+b_n' \leq C \rho_1^n.
$$
Then we conclude $|a_n'+b_n'| \to 0$ as $n \to \infty$. Exactly the
same derivation can be applied to multiple path, it follows that
$$
q_n \leq \rho^n,  \qquad  q_n' \leq C \rho_1^n.
$$
So no matter what level we started from the deleted intervals, as
long as they have long common tails, the corresponding values of
$K_2$ function are close. Therefore $K_2$ is well defined.
\end{proof}

\section{Computation of $H''(Z)|_{\varepsilon=1/2}$} \label{Low-SNR}

Let
$$
\mathbf{p}_n=[p(Z_1^n, E_n=0), p(Z_1^n, E_n=1)],
$$
and
$$
\mathbf{M}(Z_{n-1}, Z_n)=\left[ \begin{array}{cc}
                                   (1-\varepsilon) p_X(Z_n|Z_{n-1})& \varepsilon p_X(\bar{Z}_n|Z_{n-1})\\
                                   (1-\varepsilon) p_X(Z_n|\bar{Z}_{n-1})& \varepsilon
                                   p_X(\bar{Z}_n|\bar{Z}_{n-1})\\
                                \end{array}
                         \right].
$$
Then we have
$$
\mathbf{p}_n=\mathbf{p}_{n-1} \mathbf{M}(Z_{n-1}, Z_n).
$$
Immediately we obtain
$$
p_Z(Z_1^n)=\mathbf{p}_1 \mathbf{M}(Z_1, Z_2) \cdots
\mathbf{M}(Z_{n-1}, Z_n) \mathbf{1}.
$$

We consider the case when the channel is operating on the low SNR
region. For convenience, we let
$$
1-\varepsilon=\frac{1}{2}+ \delta,
$$
and
$$
\varepsilon=\frac{1}{2}- \delta.
$$
Thus when the SNR is very low, namely $\varepsilon \rightarrow
\frac{1}{2}$, correspondingly we have $\delta \rightarrow 0$. Since
$H(Z)$ is an even function at $\delta=0$, the odd order derivatives
at $\delta=0$ are all equal to $0$. In the sequel, we shall compute
the second derivative of $H(Z)$ at $\delta=0$.

In this case, we can rewrite the random matrix
$\mathbf{M}_i=\mathbf{M}(z_i z_{i+1})$ in the following way:
$$
\mathbf{M}_i=\frac{1}{2} \left[\begin{array}{cc}
                                   p_X(z_{i+1}|z_i)&p_X(\bar{z}_{i+1}|z_i)\\
                                   p_X(z_{i+1}|\bar{z}_i)&p_X(\bar{z}_{i+1}|\bar{z}_i)
                               \end{array}\right]+\delta \left[\begin{array}{cc}
                                                      p_X(z_{i+1}|z_i)&-p_X(\bar{z}_{i+1}|z_i)\\
                                                      p_X(z_{i+1}|\bar{z}_i)&-p_X(\bar{z}_{i+1}|\bar{z}_i)\\
                                                  \end{array}\right].
$$
For the special case when $i=0$, we have
$$
\mathbf{M}_0=\frac{1}{2}\left[p_X(z_1), p_X(\bar{z}_{i+1}) \right]+
\delta \left[p_X(z_1), -p_X(\bar{z}_1)\right].
$$
Then
$$
p_Z(z_1^n)=(\frac{1}{2} \mathbf{M}_0^{(0)} + \delta
\mathbf{M}_0^{(1)})(\frac{1}{2} \mathbf{M}_1^{(0)} + \delta
\mathbf{M}_1^{(1)}) \cdots (\frac{1}{2} \mathbf{M}_{n-1}^{(0)} +
\delta \mathbf{M}_{n-1}^{(1)}) \mathbf{1}.
$$
Now define the function
$$
\mathbf{R_n}(\delta)= \sum_{z_1^n} p_Z(z_1^n) \log(p_Z(z_1^n)).
$$
Then according to the definition of $H(Z)$,
$$
H(Z)=-\lim_{n \to \infty} \frac{1}{n} \mathbf{R}_n(\delta).
$$
It can be checked that
$$
\frac{\partial \mathbf{R}_n (\delta)}{\partial \delta}= \sum_{z_1^n}
\frac{\partial p_Z(z_1^n)}{\partial \delta} (\log p_Z(z_1^n)+1)).
$$
Now
$$
\left. \frac{\partial p_Z(z_1^n)}{\partial \delta}
\right|_{\delta=0}=\left(\frac{1}{2}\right)^{n-1} \sum_{i=0}^{n-1}
\mathbf{M}_0^{(0)} \mathbf{M}_1^{(0)} \cdots \mathbf{M}_{i-1}^{(0)}
\mathbf{M}_i^{(1)}\mathbf{M}_{i+1}^{(0)} \cdots
\mathbf{M}_{n-1}^{(0)} \mathbf{1}
$$
$$
=\left(\frac{1}{2}\right)^{n-1} \sum_{i=1}^n
(p_X(z_i)-p_X(\bar{z}_i)).
$$
Again simple calculations will lead to
$$
\frac{\partial^2 \mathbf{R}_n (\delta)}{\partial
\delta^2}=\sum_{z_1^n} \left( \frac{\partial^2 p_Z(z_1^n)}{\partial
\delta^2} \log p_Z(z_1^n)+\frac{1}{p_Z(z_1^n)}\left(\frac{\partial
p_Z(z_1^n)}{\partial \delta}\right)^2+\frac{\partial^2
p_Z(z_1^n)}{\partial \delta^2} \right).
$$
Since
$$
\left. \frac{\partial^2 p_Z(z_1^n)}{\partial \delta^2}
\right|_{\delta=0}=\left(\frac{1}{2}\right)^{n-2} \sum_{i \ne j}
\mathbf{M}_0^{(0)} \mathbf{M}_1^{(0)} \cdots \mathbf{M}_{i-1}^{(0)}
\mathbf{M}_i^{(1)}\mathbf{M}_{i+1}^{(0)} \cdots
\mathbf{M}_{j-1}^{(0)} \mathbf{M}_j^{(1)}\mathbf{M}_{j+1}^{(0)}
\cdots \mathbf{M}_{n-1}^{(0)} \mathbf{1}
$$
$$
=\left(\frac{1}{2}\right)^{n-2} \left[p_X(z_{i+1}),
-p_X(\bar{z}_{i+1}) \right] \left[\begin{array}{cc}
                                 p_X(z_{j+1}|z_{i+1})&-p_X(\bar{z}_{j+1}|z_{i+1})\\
                                 p_X(z_{j+1}|\bar{z}_{i+1})&-p_X(\bar{z}_{j+1}|\bar{z}_{i+1})\\
                            \end{array}\right]
$$
$$
=\left(\frac{1}{2}\right)^{n-2} \sum_{i \ne j} (p_X(z_{j+1},
z_{i+1})-p_X(z_{j+1}, \bar{z}_{i+1})-p_X(\bar{z}_{j+1},
z_{i+1})+p_X(\bar{z}_{j+1}, \bar{z}_{i+1})),
$$
we have
$$
\left. \frac{\partial^2 \mathbf{R}_n (\delta)}{\partial \delta^2}
\right|_{\delta=0} =\sum_{z_1^n} 2^n \left(
\left(\frac{1}{2}\right)^{n-1} \sum_{i=1}^n
(p_X(z_i)-p_X(\bar{z}_i))\right)^2.
$$

Let $x, y$ temporarily denote the stationary distribution
$$
p_X(0)=\frac{\pi_{10}}{\pi_{01}+\pi_{10}},  \qquad
p_X(1)=\frac{\pi_{01}}{\pi_{01}+\pi_{10}},
$$
respectively. Then
\begin{eqnarray*}
\left. \frac{\partial^2 \mathbf{R}_n (\delta) }{\partial \delta^2}
\right|_{\delta=0} &=&\frac{1}{2^{n-2}} \sum_{i=0}^n C_n^i (2 i
x+2 (n-i) y -n)^2\\
&=&\frac{1}{2^{n-2}} \sum_{i=0}^n C_n^i ( (2 x -2 y) i+ 2 n y
-n)^2\\
&=&(2 x -2 y)^2 \sum_{i=0}^n C_n^i i^2+ (2ny-n)^2 \sum_{i=0}^n 1+
2(2x-2y)(2ny-n) \sum_{i=0}^n C_n^i i.
\end{eqnarray*}

Using the following two combinatoric identity
$$
\sum_{i=0}^n i C_n^i=n 2^{n-1},
$$
and
$$
\sum_{i=0}^n i^2 C_n^i=n(n-1)2^{n-2}+n 2^{n-1},
$$
we derive
$$
\left. \frac{\partial^2 \mathbf{R}_n (\delta)}{\partial \delta^2}
\right|_{\delta=0} = \frac{1}{2^{n-2}} \left((x-y)^2 ( n(n-1) 2^n +
n 2^{n+1}) + n^2 2^n (2y-1)^2+2 (x-y)(2y-1) n^2 2^n \right)
$$
$$
=4n(x-y)^2.
$$

From the fact that the derivatives of $H(Z)$ with respect to
$\varepsilon$ are uniformly bounded on $[0, 1/2]$ (see~\cite{ja04},
also implied by Theorem 1.1 of~\cite{gm05}  and the computation of
$H^{\varepsilon}(Z)|_{\varepsilon=0}$), we draw the conclusion that
the second coefficient of $H(Z)$ is equal to
$$
H''(Z)|_{\varepsilon=1/2}=-4
\left(\frac{\pi_{10}-\pi_{01}}{\pi_{10}+\pi_{01}} \right)^2.
$$

\end{document}